\documentclass[aoas,MSNbibl,nameyear,dvips]{arximspdf}
\usepackage{multirow}
\usepackage{dcolumn}
\usepackage{graphicx}

\doi{10.1214/12-AOAS612} 
\volume{7}
\issue{2}
\pubyear{2013}
\firstpage{846}
\lastpage{859}

\makeatletter

\let\sv@tabnotetext\tabnotetext
\let\sv@tabnotemark@fmt\tabnotemark@fmt
\long\def\legend#1{{\let\tabnote@indent\leavevmode\sv@tabnotetext[]{}{#1}}}

\newcolumntype{d}[1]{D{.}{.}{#1}}

\newcommand{\boldn}{\mathbf{n}}
\newcommand{\boldr}{\mathbf{r}}

\makeatother

\begin{document}
\begin{frontmatter}

\title{The number of killings in southern rural Norway,~1300--1569}
\runtitle{Killings in Norway}

\begin{aug}
\author[A]{\fnms{Joseph B.} \snm{Kadane}\ead[label=e1]{kadane@stat.cmu.edu}}
\and
\author[B]{\fnms{Ferdinand L.} \snm{N\ae shagen}\corref{}\ead[label=e2]{linaesha@online.no}}
\runauthor{J. B. Kadane and F. L. N\ae shagen}
\affiliation{Carnegie Mellon University and The Norwegian Police
University College}
\address[A]{Department of Statistics\\
Carnegie Mellon University\\
Pittsburgh, Pennsylvania 15213\\
USA\\
\printead{e1}}
\address[B]{The Research Department\\
The Norwegian Police University College\\
Oslo\\
Norway\\
\printead{e2}} 
\end{aug}

\received{\smonth{5} \syear{2012}}
\revised{\smonth{10} \syear{2012}}

%
\begin{abstract}
Three dual systems estimates are employed to study the number of
killings in southern rural Norway in a period of slightly over 250
years. The first system is a set of five letters sent to each killer as
part of the legal process. The second system is the mention of killings
from all other contemporary sources. The posterior distributions
derived suggest fewer such killings than rough demographic estimates.
\end{abstract}

%

\end{frontmatter}

\section{Norwegian homicide law and the documentary evidence}\label{sec2}

This paper studies the number of
killings in Norway in the period 1300--1569, that is, the last fifty
years of Norway's High Middle Age, through the Late Middle Ages, and a
generation or so into the Early Modern Age. The extant written data
about such killings, is of course, only a fraction of the documents issued.

Certain homicides (and some other crimes) were ``noncompensation
crimes'' (\textit{ubotemal}), which means that they, unless the king
decided otherwise, were atoned for by capital punishment or outlawry
and confiscation of the criminal's property. Noncompensation homicides
would, for instance, be the killing of a man in his own house, the
killing of a kinsman, or a killing on a holy day. A study of the
documents issued in such cases shows that King Magnus the Lawmender's
National Law of 1274 was systematically set aside in such cases, for
good economic reasons. There would be no compensation to the victim's
next of kin, and it might even be a loss to the king's district officer
(\textit{sysselmann}, the equivalent of an English sheriff)
if he had to pay an executioner the equivalent of a
craftsman's monthly pay for decapitating a pennyless youngster. With,
however, an economic atonement for the killing (\textit{botemal}), the
vicim's heirs would get their compensation, and the king's district
officer would get the fine [strictly speaking, two fines, a recently
introduced one for depriving the king of a subject (\textit{tegngilde})
and an older one for the king's pardon (\textit{fredkjop}), similar to the
continental Germanic \textit{fredus}] nominally due to the king, which was
about fifty percent of the normal compensation. In case of
noncompensation killings the fine would be relatively higher, one
regular fine for a killing, to which would be added another one for the
killing of a brother, a second if it took place in his own house, and a
third if it took place on a holy day. As we can see from some
documents, family members would help to pay even though their legal
obligation to do so had been abolished in 1260. The loss of a family
member, cherished or not, would weaken the family. Some may have
contributed in money or species, others may have guaranteed as
securities as some documents show. Furthermore, there was some
opportunity for haggling and the period before the compensation or fine
was fully paid might on occasion be considerably longer than the year
specified in the letter of pardon.

This process had five documents as its outcome. The killer, who was
left at large and indeed might be said to be the prosecutor, had first
to go to the King's Chancellor in Oslo to get a protection letter
(\textit{gridsbrev}) which both gave him a temporary protection against avengers
and also was an order to the king's district officer to hear the case
so as to find whether the killer had fulfilled the obligation of taking
public responsibility for the killing and also whether he had sureties
for the payment of compensation and fine. In accordance with this the
district officer held a hearing with witnesses and the parties present
and issued an evidence letter (\textit{provsbrev}) summing up the relevant
facts, including what might make this one or several ubotemal. With
this provsbrev the killer had once more to travel to the King's
Chancellor who then issued a permanent pardon (\textit{landsvist}, right
to stay in the country) which also stated the amount to be paid in
fine, and the condition that compensation and fine were to be paid
within a year. As we can see, practice did at times give the killer
several years respite before these sums were paid, but when paid they
resulted in one receipt from the king's district officer and one from
the victim's heirs. These five letters were all preserved in the
killer's archive as part of a farm archive together with deeds,
inheritance divisions etc. until fire, wetness or some overly tidy
daughter-in-law put an end to the existence of the large majority.

Supplementary material [\citet{suppA}]
is an index of the documents that did survive, showing evidence of 337
killings in this time period. Of these, 194 are documented from the
killer's archive, 143 are only from other sources and 4 are mentioned
both in the killer's archive and in other sources. The other sources
are quite varied, but include local officials, the King's Chancellor,
regional potentates, church officials, and private letters and diaries.
The data used in this paper are summarized in Table \ref{tab1}.

%
\begin{table}
\caption{Two-way classification of records of killings}\label{tab1}
\begin{tabular*}{\tablewidth}{@{\extracolsep{4in minus 4in}}lccd{3.0}d{2.0}cccc@{}}
\hline
& & \multicolumn{6}{c}{\textbf{Number of letters from killer's archive}} &\\[-4pt]
& & \multicolumn{6}{l}{\hrulefill} &\\
& & \multicolumn{1}{c}{\textbf{0}} & \multicolumn{1}{c}{\textbf{1}}
& \multicolumn{1}{c}{\textbf{2}} & \multicolumn{1}{c}{\textbf{3}} & \multicolumn{1}{c}{\textbf{4}}
& \multicolumn{1}{c}{\textbf{5}} & \textbf{Total}\\
\hline
\multirow{2}{*}{Mentioned in other sources?} & No & $n$ & 162 & 20 & 5 & 3 & 0 & $190 +
n$\\
& Yes & 143 & 3 & 0 & 1 & 0 & 0 & 147\\ [6pt]
Total & & $143+n$ & 165 & 20 & 6 & 3 & 0 & $337+n$\\
\hline
\end{tabular*}
\end{table}



The purpose is to find a distribution for $n$ and hence for $337 + n$,
the total number of killings in the period.

\section{Demographic evidence about the number of killings}\label{secdemo}

During this period Norway (like other European countries) underwent
dramatic demographic changes. There is, furthermore, some disagreement
about absolute numbers in given years during this period, but the most
recent text book authors agree that when the plague first hit Norway in
1349 its population may have been 500,000 and perhaps slightly lower in
the preceding half century. The recurrent plague epidemics reduced the
population to its lowest point ca. 1450 to 1500, ca 200,000 or perhaps
less [\citet{moseng2007}, pages 233--236, 294 and 295]. After this population
started growing again and, in spite of recurrent epidemics, grew to
440,000 in the 1660s, the first really reliable assessment. These
estimates concern Norway as it was then, before the country had lost
almost ten percent of its territory and population due to Danish
military misadventures. The data used here are, for the sake of
comparison, only taken from present-day Norwegian territory, so about
ten percent should be deducted from population estimates.

With two exceptions there is no conspicuous geographic bias in the
data. Telemark, which both in the Middle Ages and later had a
reputation for violence, is very well represented in these data. Due to
the cases where the scene of the homicide is geographically localized,
or that of the person paying for receiving compensation or fine, or
their provenience (come to an archive from a rural district) is, and
the fact that family archives are preserved in rural districts, as farm
archives while similar urban archives are unknown, we can be fairly
sure that scarcely any of these documents had an urban origin---which
means that they reflect the situation in the countryside, not in the
much more violent cities and towns. This may account for the
discrepancy between the homicide estimates for the mid-sixteenth
century (10--15 per 100,000) made from another type of data (accounts of
fines and confiscations) by \citet{naeshagen2005}, and the somewhat
lower estimates this study yields. Only about 3 percent of the
population lived in the three larger cities, Bergen, Trondheim and
Oslo, but their population showed an extreme inclination to homicide.
Thus, Bergen, Norway's largest and most heterogenous city, with a
population of 6,000 had from 1562 to 1571 a homicide rate of 83 per
100,000 [\citet{sandnes1990}, pages 72--74]. Thus, with these rural data
one should expect a somewhat lower estimate than N{\ae}shagen's 10 to 15
per 100,000 from the mid-sixteenth century which includes cities (2005).

Central Norway (Tr$\o$ndelag) and Northern Norway with, respectively, 13
and 11 percent of the population [\citet{dyrvik1979}, page 18] seem not
to be represented among
these documents. Judging from the mid-sixteenth-century lists of fines
and confiscations, homicides may have been rarer in Central Norway than
in the rest of the country, while Northern Norway does not distinguish
itself in any way [\citet{naeshagen2005}, page 416], and later data
support the conclusion about Central Norway [\citet{sandnes1990}, page 79].

So supposing that the population of Norway as it was then was 500,000
in the period from 1300 to 1350, and roughly 200,000 in the period from
1350 to 1569, we must deduct 10\% to account for the territory lost.
This yields 450,000 in 1300 to 1350, and 180,000 for the later period.
Additionally, we deduct 24\% (13\% in Central Norway, 11\% in Northern
Norway) for rural areas not covered, and another 3\% for the cities,
yielding a deduction of 27\%. Thus, we estimate rural southern Norway
to have had a population of 330,000 in the period from 1300 to 1350,
and 130,000 from 1350 to 1569. It should be emphasized that these are
rough estimates only.

The next set of estimates concerns the rate of killings. Accepting the
estimates from somewhat later of 10 to 15 per hundred thousand per year
overall, but a much higher rate (83 per hundred thousand) for the 3\%
of the urban population suggests a rate of 8 to 13 per hundred thousand
per year in rural southern Norway.

Applied to the 50 year period before the plague and the 219 years after
the plague, this yields a range of 3600 to 5850 for the number of
killings in rural southern Norway during the period in question.

\section{Models of the data}\label{secmodels}

Problems of missing data are ubiquitous; indeed, every parameter not
known with certainty can be regarded as ``missing data'' in some
sense. In biostatistics, survival analysis can be regarded as a method
for dealing with missing time-of-death data for patients still
alive. But these problems are especially acute in history, geology,
the interpretation of fossils, astronomy and archeology. In one
instance, \citet{kadanehastorf1988}, the authors assumed known
preservation probabilities for different kinds of burnt seeds in an
archeological site in Peru.

While the methods used here bear a relationship with problems of
estimating the number of species [see \citet{bungefitzpatrick1993} for
a review], the more closely related literature is that of dual systems
estimators, growing out of the early work of \citet{petersen1896} and
\citet{lincoln1930}, and applied to the problem of census coverage by
\citet{wolter1986}.

\subsection*{A. Simple dual systems}
The simplest treatment of data of this kind is to amalgamate all
mentions in the killer's archive together, resulting in the following
$2 \times2$ table.

%
\begin{table}
\caption{Reduced data}\label{tab2}
\begin{tabular*}{295pt}{@{\extracolsep{4in minus 4in}}llcrc@{}}
\hline
& & \multicolumn{2}{c}{\textbf{Killer's archive?}}\\[-4pt]
& & \multicolumn{2}{c}{\hrulefill}\\
& & \textbf{No} & \multicolumn{1}{c}{\textbf{Yes}} & \textbf{Total}\\
\hline
\multirow{2}{*}{Mentioned in other sources?} & No & $n$ & 190 & $190+n$\\
& Yes & 143 & 4 & 147\\ [6pt]
Total & & $143+n$ & 194 & $337+n$\\
\hline
\end{tabular*}
\end{table}

To establish notation for this case, let the numbers in Table
\ref{tab2} be represented as shown in Table \ref{tab3}.

%
\begin{table}[b]
\tablewidth=295pt
\caption{General notation for Table \protect\ref{tab2}} \label{tab3}
\begin{tabular*}{\tablewidth}{@{\extracolsep{4in minus 4in}}llccc@{}}
\hline
& & \multicolumn{2}{c}{\textbf{Killer's archive?}}\\[-4pt]
& & \multicolumn{2}{c}{\hrulefill}\\
& & \textbf{No} & \multicolumn{1}{c}{\textbf{Yes}} & \textbf{Total}\\
\hline
\multirow{2}{*}{Mentioned in other sources?} & No & $n_{00}$ & $n_{01}$ & $n_{0+}$\\
& Yes & $n_{10}$ & $n_{11}$ & $n_{1+}$\\ [6pt]
Total & & $n_{+0}$ & $n_{+1}$ & $n_{++}$\\
\hline
\end{tabular*}
\legend{Note: $n_{00} = n$.}
\end{table}

The data can be taken to be multinomial, with probabilities $p_{ij}$,
and hence likelihood
%
\begin{equation}
\label{eq1} L = \pmatrix{n_{++}
\cr
n_{00},
n_{01}, n_{10}, n_{11}} \mathop{\prod
_{i=0,1}}_{j=0,1}p^{n_{ij}}_{ij}.
\end{equation}

A key assumption is that of independence, which would mean that whether
a killing is known from the preservation of a letter from the killer's
archive has no bearing on whether it is known from the other sources.
In this application, such an assumption seems entirely reasonable. So
if $p$ is the probability a killing is mentioned in other sources and
$q$ is the probability a killing is known from at least one letter from
the killer's archive, the assumption of independence can be written as
%
\begin{equation}
\label{eq2} p_{ij} = p^{i}\overline{p}{}^{\overline{i}}q^{j}
\overline {q}{}^{\overline{j}},\qquad i=0,1; j=0,1,
\end{equation}
where $\overline{x} = 1-x$.

Substituting (\ref{eq2}) into (\ref{eq1}) yields
%
\begin{equation}
\label{eq3} L = \pmatrix{n_{++}
\cr
n_{00},
n_{01},n_{10},n_{11}} p^{n_{1+}}
\overline{p}{}^{n_{0+}}q^{n_{+1}} \overline{q}{}^{n_{+0}}.
\end{equation}

The parameters $p, q$ and $n$ are all that matter here, and $n$ is the
parameter of interest. Any reasonable prior distribution (i.e., one
that is not strongly opinionated) for $p$ and $q$ will lead to the same
inference, given the values of $n_{0+},n_{1+},n_{+0}$ and $n_{+1}$ in
this data set. Hence, we accept independent uniform priors for $p$
and~$q$. In view of the material in Section \ref{secdemo}, the prior of
interest on the total number of killings, $n+337$, is uniform $(337,
5850)$. However, for the first computation reported here we use a much
broader uniform prior on $n$ in order to show the uncertainty inherent
in the likelihood.

\begin{figure}

\includegraphics{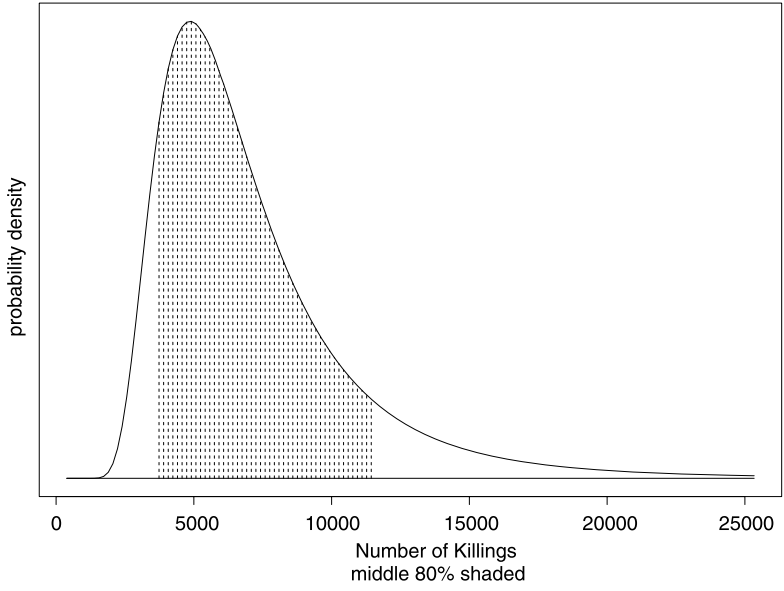}

\caption{Simple dual systems integrated likelihood.}\label{fig1}
\end{figure}

Using the well-known integration result,
%
\begin{eqnarray}
\label{eq4}
\int^{1}_{0}x^{n}(1-x)^{m}
\,dx &=& B(n+1, m+1) = \frac{\Gamma
(n+1)\Gamma(m+1)}{\Gamma(n+m+2)} \nonumber\\[-8pt]\\[-8pt]
&=& \frac{n!m!}{(n+m+1)!},\nonumber
\end{eqnarray}
the integrated likelihood is
%
\begin{equation}
\label{eq5} \pmatrix{n_{++}
\cr
n_{00},n_{01},n_{10},n_{11}}
\frac{n_{1+}!
n_{0+}!n_{+1}!n_{+0}!}{[(n_{++}+1)!]^{2}}.
\end{equation}
Now $n_{01}, n_{10},n_{11},n_{+1}$ and $n_{1+}$ do not depend on $n$.
Hence, these factors do not matter for the integrated likelihood,
yielding an integrated likelihood proportional to
%
\begin{equation}
\label{eq6} \frac{(n_{0+})!(n_{+0})!}{n_{00}!(n_{++}+1)(n_{++}+1)!} = \frac{(n+190)! (n+143)!}{n!(n+338)(n+338)!}.
\end{equation}
Figure \ref{fig1} plots, as a probability distribution, the quantity
$n+337$, the total number of killings. Implicitly the prior on $n$ used
in this calculation is uniform with an upper bound of at least 25,000,
which is much higher than we find credible. Nonetheless, for display
purposes, we show it.


The quantiles of the data in Figure \ref{fig1} are reported in
Table \ref{tab4}. Together Figure~\ref{fig1} and Table \ref{tab4}
suggest substantial uncertainty about the total number of killings; the
middle 80\% of the distribution lies between 3337 and 10,837, a gap of
7500 killings; the median of the distribution is 5837.

%
\begin{table}[b]
\caption{Quantiles for Figure \protect\ref{fig1}}\label{tab4}
\begin{tabular*}{\tablewidth}{@{\extracolsep{\fill}}lccccccccc@{}}
\hline
Quantile & 3337 & 3837 & 4337 & 4837 & 5837 & 6337 & 7337 & 8337 &
10,837\\
Probability & 0.1 & 0.2 & 0.3 & 0.4 & 0.5 & 0.6 & 0.7 & 0.8 & 0.9\\
\hline
\end{tabular*}
\end{table}

This suggests the desirability of making more use of the data in
Table \ref{tab1}, and in particular the data on the number of letters
found in each killer's archive.

\subsection*{B. Dual systems binomial model}

To do so, we now establish general notation for Table \ref{tab1}, in
Table \ref{tab5}.

Let $\boldn= (n_{00}, n_{01}, n_{02}, \ldots, n_{05}, n_{10},
n_{11},\ldots, n_{15})$ and $\boldn! = \prod^{5}_{i=0} \prod^{1}_{j=0}
n_{ij}!$.

Then the multinomial likelihood can be written as
%
\begin{equation}
\label{eq7} L = \frac{n_{++}!}{\boldn!} \mathop{\prod_{i=0,1}}_{j=0,\ldots,5}p^{n_{ij}}_{ij}.
\end{equation}
Again imposing independence, we have
%
\begin{equation}
\label{eq8} p_{ij} = r_{j}s^{i}
\overline{s}{}^{\overline{i}},\qquad j=0,\ldots, 5; i=0,1,
\end{equation}
where $r_{j}$ is the probability of $j$ surviving letters in the
archive and $s$ is the probability of being mentioned in other sources.

%
\begin{table}
\caption{Notation for Table \protect\ref{tab1}} \label{tab5}
\begin{tabular*}{\tablewidth}{@{\extracolsep{4in minus 4in}}lcccccccc@{}}
\hline
& & \multicolumn{6}{c}{\textbf{Number of letters in killer's archive}} &\\[-4pt]
& & \multicolumn{6}{l}{\hrulefill} &\\
& & \multicolumn{1}{c}{\textbf{0}} & \multicolumn{1}{c}{\textbf{1}}
& \multicolumn{1}{c}{\textbf{2}} & \multicolumn{1}{c}{\textbf{3}} & \multicolumn{1}{c}{\textbf{4}}
& \multicolumn{1}{c}{\textbf{5}} & \textbf{Total}\\
\hline
\multirow{2}{*}{Mentioned in other sources?} & No & $n_{00}$ & $n_{01}$ & $n_{02}$ &
$n_{03}$ & $n_{04}$ & $n_{05}$ & $n_{0+}$\\
& Yes & $n_{10}$ & $n_{11}$ & $n_{12}$ & $n_{13}$ & $n_{14}$ & $n_{15}$
& $n_{1+}$\\ [6pt]
Total & & $n_{+0}$ & $n_{+1}$ & $n_{+2}$ & $n_{+3}$ & $n_{+4}$ &
$n_{+5}$ & $n_{++}$\\
\hline
\end{tabular*}
\end{table}

Substituting (\ref{eq8}) into (\ref{eq7}), we obtain
%
\begin{equation}
\label{eq9} L= \frac{n_{++}!}{\boldn!} \prod^{5}_{j=0}
r^{n+j}_{j} s^{n_{1+}}\overline{s}{}^{n_{0+}}.
\end{equation}

A simple model to impose on $\boldr= (r_{o}, r_{1},\ldots, r_{5})$ is
binomial $(5,p)$, where $p$ is here the probability that each letter in
a killer's archive survives (this assumption is revisited in
subsection~C, ahead). With the binomial assumption,
%
\begin{equation}
\label{eq10} r_{j} = \pmatrix{5
\cr
j} p^{j}
\overline{p}{}^{5-j},\qquad j=0,\ldots,5.
\end{equation}

Then
%
\begin{equation}
\label{eq11} \prod^{5}_{j=0}
r^{n+j}_{j} = \prod^{5}_{j=0}
\pmatrix{5
\cr
j,5-j}^{n_{+j}} p^{\sum^{5}_{j=0}jn_{+j}}\overline{p}{}^{\sum
^{5}_{j=0}(5-j)n_{+j}}.
\end{equation}
Let $S_{1} = \sum^{5}_{j=0}j n_{+j}$. Then $\sum^{5}_{j=0}(5-j)n_{+j}
= 5n_{++}-S_{1}$.

Hence,
%
\begin{equation}
\label{eq12} \prod^{5}_{j=0}r_{j}^{n_{+j}}=
\prod^{5}_{j=0} \pmatrix{5
\cr
j,5-j}^{n_{+j}}p^{S_{1}}\overline{p}{}^{5n_{++}-S_{1}}.
\end{equation}

The first term on the right can be written for our data as
%
\begin{eqnarray}
\label{eq13}\quad
&&\prod^{5}_{j=0} \pmatrix{5
\cr
j,5-j}^{n_{+j}}\nonumber\\[-8pt]\\[-8pt]
&&\qquad= \biggl(\frac
{5!}{0!5!} \biggr)^{n+143} \biggl(
\frac{5!}{1!4!} \biggr)^{165} \biggl(\frac{5!}{2!3!}
\biggr)^{20} \biggl(\frac{5!}{3!2!} \biggr)^{6} \biggl(
\frac{5!}{4!1!} \biggr)^{3} \biggl(\frac{5!}{0!5!}
\biggr)^{0}.\nonumber
\end{eqnarray}
Only the first term has an exponent that depends on a parameter, and
that term is 1 raised to a power, so the entire product is constant
with respect to the parameters, and can be dropped. Similarly, in the
terms for $\boldn!$ only the first, $n !$, depends on the parameters,
and the others can be dropped:
%
\begin{equation}
\label{eq14} L \propto\frac{(n_{++})!}{n!} p^{S_{1}} \overline{p}{}^{5n_{++}-S_{1}}
s^{n_{1+}}\overline{s}{}^{n_{0+}}.
\end{equation}
Again, using (\ref{eq4}) and independent uniform distributions on $p$
and $s$, the integrated likelihood for $n$ is
%
\begin{eqnarray}
\label{eq15}
&&
\frac{(n_{++})!}{n!} \frac{(S_{1})!
(5n_{++}-S_{1})!}{(5n_{++}+1)!} \frac{(n_{1+})!(n_{0+})!}{(n_{++}+1)!}
\nonumber\\[-8pt]\\[-8pt]
&&\qquad= \frac{S_{1}!(5n_{++}-S_{1})! (n_{1+})!
(n_{0+})!}{n!(5n_{++}+1)!(n_{++}+1)}.\nonumber
\end{eqnarray}

Finally, $S_{1}$ and $n_{1+}$ also do not depend on $n$, so those terms
can be dropped as well, yielding the integrated likelihood proportional to
%
\begin{equation}
\label{eq16} \frac{(5n_{++}-S_{1})! (n_{0+})!}{n!(5n_{++}+1)! (n_{++}+1)}.
\end{equation}

\begin{figure}

\includegraphics{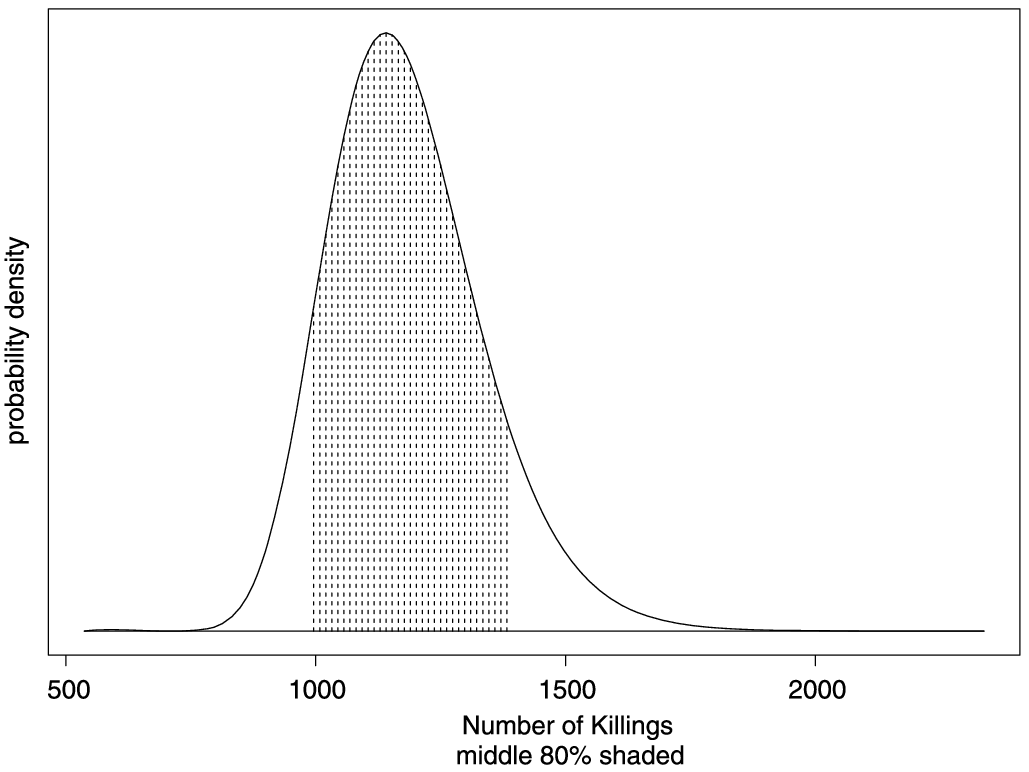}

\caption{Binomial dual systems posterior distribution. \textup{Note that
Figure \protect\ref{fig1} has a wider scale of the number of killings.}}\label{fig2}
\end{figure}

Figure \ref{fig2} plots the posterior distribution for $n+337$ whose
quantiles are given in Table \ref{tab6}. Here the median is
1155.


%
\begin{table}[b]
\caption{Quantiles for dual systems posterior distribution under the
binomial model}\label{tab6}
\begin{tabular*}{\tablewidth}{@{\extracolsep{\fill}}lccccccccc@{}}
\hline
Quantile & 978 & 1037 & 1076 & 1116 & 1155 & 1195 & 1234 & 1293 & 1372\\
Probability & 0.1 & 0.2 & 0.3 & 0.4 & 0.5 & 0.6 & 0.7 & 0.8 & 0.9\\
\hline
\end{tabular*}
\end{table}

Thus, this model suggests remarkably fewer killings than those
suggested by the simple dual systems estimate reported in Figure
\ref{fig1} and Table \ref{tab6}.

\subsection*{C. Com-binomial model}

The binomial model implies that the survival of a document from a
killer's archive is an event independent of the survival of other
documents from the same killer's archive. Since all five letters are
addressed to the same person (the killer), it is likely that they would
tend to be stored together. Hence, it seems prudent to expand the model
to allow for positive correlation among the events of survival of
letters addressed to the same killer. [A referee suggests that an
overly tidy daughter-in-law may have kept only one letter, leading to
negative correlation. While that may have happened in a few instances,
we think that joint physical destruction (fire and water) is far more
likely, and hence expect positive correlation in the survival event of
documents from a killer's archive.]

One model that allows for such correlation is the com-binomial
distribution [\citet{shmueli-etal2005}]. The pdf for this distribution
is given by
%
\begin{equation}
\label{eq17}\quad P\{X=j|p,\nu\} = \frac{p^{j}(1-p)^{m-j}
{m\choose j,m-j}^{\nu}}{\sum^{m}_{k=0} p^{k}(1-p)^{m-k} {m\choose k,
m-k}^{\nu}},\qquad j = 0, 1,\ldots, m.
\end{equation}
When $\nu=1$, this distribution reduces to the binomial distribution,
and hence to independence of survival of the documents sent to a given
killer. For $\nu> 1$, the survival would be negatively correlated. For
$\nu< 1$, the survival would be positively correlated. In this
application, the latter is expected. As $\nu\rightarrow\infty$, the
probability would become concentrated on a single point. As $\nu
\rightarrow- \infty$, it would become concentrated on 0 and $m$.

\begin{figure}

\includegraphics{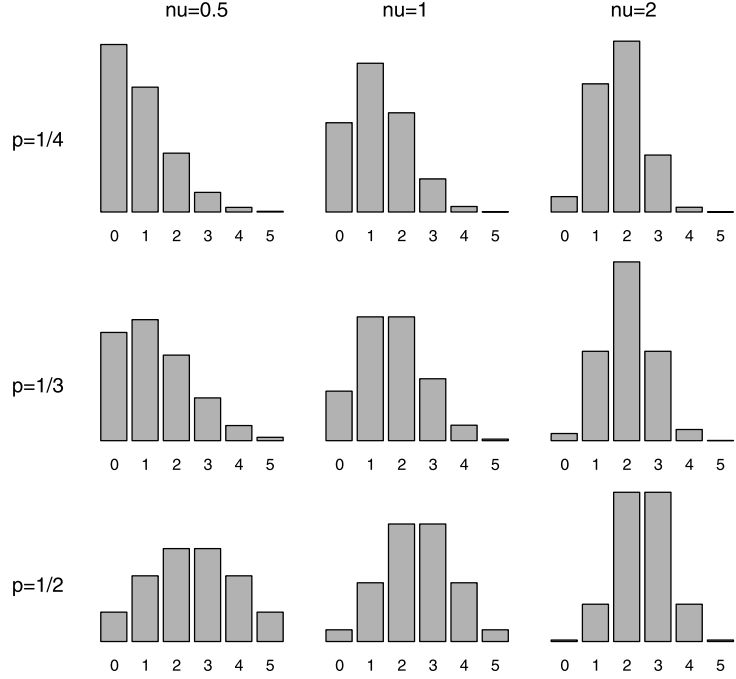}

\caption{Com-binomial distribution for various values of $p$ and
$nu$.}\label{fig3}
\end{figure}

Because this distribution is unfamiliar, it is perhaps useful to look
at some examples, displayed in Figure \ref{fig3} for the case $m=5$,
which is the value of $m$ in this application. In this figure, looking
across rows, as $\nu$ increases, the probability tends to concentrate
on a single point (except at $p= 1/2$, where symmetry leads to two
dominant points, 2 and 3).


As alluded to above, values of $\nu$ above 1 do not make sense in this
application. Therefore, the analysis to be presented imposes the
condition $\nu\leq1$ as a hard constraint, by using a prior that put
zero probability in the space $\nu> 1$.

To incorporate the com-binomial distribution into the model, $r_{j}$ in
(\ref{eq10}) is replaced by the expression in (\ref{eq17}). This
yields the likelihood
%
\begin{eqnarray}
\label{eq18} L & = & \frac{n_{++}!}{\boldn!} s^{n_{1+}} \overline{s}{}^{n_{0+}}
\prod^{5}_{j=0} r_{j}^{n_{+j}}
\nonumber\\[-8pt]\\[-8pt]
& = & \frac{n_{++}!}{\boldn!} s^{n_{1+}}\overline{s}{}^{n_{0+}}\prod
^{5}_{j=0} \biggl[ \frac{p^{j}(1-p)^{m-j}{m\choose j,m-j}^{\nu}} {
\sum^{m}_{k=0} p^{k}(1-p)^{m-k} {m\choose k,m-k}^{\nu}}
\biggr]^{n_{+j}}.
\nonumber
\end{eqnarray}

It is convenient to divide the numerator and denominator in the product
term by the factor $(1-p)^{m}(m !)^{\nu}$, yielding
%
\begin{equation}
\label{eq19} \frac{p^{j}(1-p)^{m-j}{m\choose j,m-j}^{\nu}}{\sum^{m}_{k=0}
p^{k}(1-p)^{m-k}{m\choose k,m-k}^{\nu}} = \frac{\theta^{j}/[j!(m-j)!]^{\nu}}{\sum^{5}_{k=0} \theta^{k}/[k
!(m-k)!]^{\nu}},
\end{equation}
where $\theta= p/(1-p)$.

It is further convenient to rewrite (\ref{eq19}) as follows:
%
\begin{eqnarray}
\label{eq20} && \theta^{j}\bigg/ \Biggl\{ \bigl[j!(m-j)!
\bigr]^{\nu} \Biggl(\sum^{5}_{k=0}
\theta ^{k}/ \bigl[k!(m-k)!\bigr]^{\nu} \Biggr) \Biggr\}
\nonumber
\\
&&\qquad= e^{j \log\theta-\nu\log[j!(m-j)!]}/Z(\theta,\nu)
\\
&&\eqntext{\mbox{where }\displaystyle  Z(\theta,\nu) = \sum^{5}_{k=0}
\theta^{k}/\bigl[k!(m-k)!\bigr]^{\nu}.}
\end{eqnarray}
Substituting (\ref{eq20}) into (\ref{eq18}) yields
%
\begin{equation}
\label{eq21} L = \frac{n_{++}!}{\boldn!} s^{n_{1+}}\overline{s}{}^{n_{0+}}
e^{s_{1}\log\theta- s_{2}\nu}/\bigl(Z(\theta, \nu)\bigr)^{n_{++}},
\end{equation}
where $s_{1}= \sum^{5}_{j=1}j n_{+j}$
and $s_{2}= \sum^{5}_{j=0} n_{+j} \log(j!(5-j)!)$.\eject

Once again $s$ can be integrated with respect to a uniform prior,
yielding the integrated likelihood
%
\begin{equation}
\label{eq22} \frac{n_{++}!}{\boldn!} \frac{(n_{1+})!(n_{0+})!}{(n_{++}+1)!} e^{s_{1}\log\theta- s_{2}\nu}/Z(\theta,
\nu)^{n_{++}}.
\end{equation}
Finally, factors not involving $\theta, \nu$ and $n$ can be
eliminated, yielding
%
\begin{equation}
\label{eq23} \frac{(n_{0+})!}{n!(n_{++}+1)} e^{s_{1}\log\theta- s_{2}\nu} Z(\theta, \nu)^{-n_{++}}.
\end{equation}

In order to have results comparable to those in Figure \ref{fig2},
proper account must be taken of the transformation from $p$ to $\theta
$. The differentials are related by
%
\begin{equation}
\label{eq24} dp = \frac{d\theta}{(1+\theta)^{2}},
\end{equation}
so $p$ uniform on $(0,1)$ is equivalent to $\theta$ having the density
$1/(1+\theta)^{2}$ on $(0,\infty)$. Thus, the form of likelihood used
here is (\ref{eq23}) multiplied by (\ref{eq24}), that is,
%
\begin{equation}
\label{eq25} \frac{(n_{0+})!}{(n_{++}+1) n!} e^{s_{1}\log\theta- s_{2}\nu} \frac
{Z(\theta,\nu)^{-n_{++}}}{(1+\theta)^{2}}.
\end{equation}
Using a grid method to integrate (\ref{eq25}) with respect to $\theta
$ and $\nu$ yields the posterior distribution in Figure \ref{fig4},
with quantiles given in Table \ref{tab7}. The median for this model
is 1143, about the same as for the binomial model.

The results of the com-binomial in Figure \ref{fig4} are very similar
to those of the binomial in Figure \ref{fig2}. The reason for this is
that the likelihood for $\nu$ strongly indicates a preference for $\nu
=1$. Glancing back at the data in Table \ref{tab1}, the data are
strongly piled up at 0 and $1$ letters from a killer's archive; there
are no killings at all for which all five letters have survived.
Therefore, the data looks much more like it would at $\nu= \infty$,
which makes no substantive sense in this problem. Given that the hard
constraint $\nu\leq1$ has been imposed, the integrated posterior puts
most weight on the largest $\nu$ permitted, that is, $\nu= 1$; the
results therefore resemble those of the binomial model reported in
Figure \ref{fig2}. While the generalization afforded by the
com-binomial did not lead to a substantially different integrated
likelihood, it was important to see whether positive correlation in the
survival of letters sent to the killer was a dominant feature of the
data. This turned out not to be the case.

\section{Conclusion}

An assumption underlying our model is that every killing resulted in
the five letters being sent to the killer. It is possible that this
is not true, and possible that the propensity to send the requisite
letters varied by geography. It is also possible that some geographical
areas were more prone to document destruction by fire, flood, etc., and
such areas might be those less carefully administered. We leave these
possibilities for further exploration.

\begin{figure}

\includegraphics{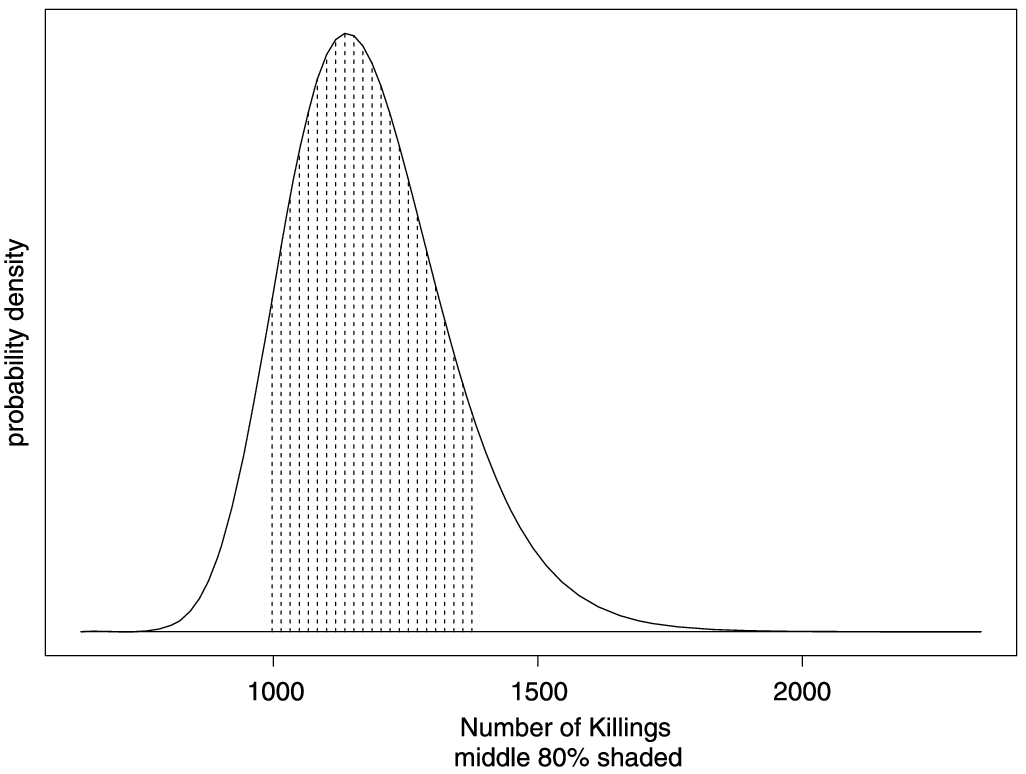}

\caption{Com-binomial posterior distribution. \textup{Note that Figure \protect\ref{fig1}
has a wider scale for the number of killings.}}\label{fig4}
\end{figure}

This paper presents three analyses of the number of killings in rural
Norway during the period in question. The first (Table \ref{tab4} and
Figure \ref{fig1}) used only the presence or absence of a mention in
the killer's archive, and found huge uncertainty in the number of
killings. The latter two, reported, respectively, in Table \ref{tab6}
and Figure~\ref{fig2}, and in Table \ref{tab7} and Figure \ref
{fig4}, are so similar that substantively they are the same. The
distribution reported indicates that perhaps rural Norway was more
peaceful in this period than had previously been thought.

%
\begin{table}[b]
\caption{Quantiles for dual systems integrated likelihood under the
com-binomial model}\label{tab7}
\begin{tabular*}{\tablewidth}{@{\extracolsep{\fill}}lccccccccc@{}}
\hline
Quantile & 959 & 1021 & 1051 & 1113 & 1143 & 1174 & 1235 & 1265 & 1357\\
Probability & 0.1 & 0.2 & 0.3 & 0.4 & 0.5 & 0.6 & 0.7 & 0.8 & 0.9\\
\hline
\end{tabular*}
\end{table}

\section*{Acknowledgments}

The authors thank their good friend Baruch Fischoff for introducing
them and suggesting that this problem might interest us both. Sarah
Brockwell did much to clean the data, and Anthony Brockwell helped with
the data structure. Jong Soo Lee also contributed to the data handling.
Conversations with Rebecca Nugent, Howard Seltman and Andrew Thomas
about \texttt{R} were also very helpful. A referee was very helpful in
correcting our rough demographic estimates of the numbers of killings.

\begin{supplement}
\stitle{Criminal homicides in Norwegian letters 1300 to 1569\\}
\slink[doi]{10.1214/12-AOAS612SUPP} 
\sdatatype{.pdf}
\sfilename{aoas612\_supp.pdf}
\sdescription{A list of letters found in Norway concerning killings
during the period of 1300 to 1569.}
\end{supplement}


\printaddresses

\end{document}